\documentclass[11pt,epsfig]{article}

\usepackage{amsmath,amssymb}
\usepackage[dvips]{graphicx,color}
\def\thefootnote{\fnsymbol{footnote}}

\newcommand{\nn}{\nonumber}
\newcommand{\SM}{{\text{SM}}}

\newcommand{\diag}{{\mathrm{diag}}}
\newcommand{\eq}{{\mathrm{eq}}}
\newcommand{\obs}{{\mathrm{obs}}}
\newcommand{\meV}{{\text{meV}}}

\newcommand{\GeV}{{\text{GeV}}}

\newcommand{\del}{\partial}
\newcommand{\deriv}[2]{\frac{\del #1}{\del #2}}
\newcommand{\Ls}{\left(}
\newcommand{\Rs}{\right)}
\newcommand{\Lm}{\left\{}
\newcommand{\Rm}{\right\}}
\newcommand{\Ll}{\left[}
\newcommand{\Rl}{\right]}

\newcommand{\RR}{\right.}
\newcommand{\PareS}[1]{\Ls{#1}\Rs}
\newcommand{\PareM}[1]{\Lm{#1}\Rm}
\newcommand{\PareL}[1]{\Ll{#1}\Rl}
\newcommand{\abs}[1]{\left|#1\right|}

\newcommand{\order}[1]{{\cal O}\Ls #1\Rs} 
\newcommand{\half}{{1\over2}}
\newcommand{\II}{I$\!$I}

\begin{document}

\begin{titlepage}
\begin{center}
{\Large\bf A novel washout effect in the flavored leptogenesis}
\end{center}
\vspace{1cm}
\begin{center}
{\bf Tetsuo Shindou}$^{}$
\footnote{E-mail: shindou@sissa.it}
and
{\bf Toshifumi Yamashita}$^{}$
\footnote{E-mail: yamasita@sissa.it}
\end{center}
\vspace{0.2cm}
\begin{center}
{\em
Scuola Internazionale Superiore di Studi Avanzati, and \\}
{\em 
Istituto Nazionale di Fisica Nucleare, I-34014 Trieste, Italy\\
}
\end{center}
\vspace{1cm}
\begin{abstract}
We investigate a flavored washout effect due to the decay of the lightest 
right-handed neutrino, 
assuming that there is non-vanishing initial lepton asymmetry
and the decay of the lightest right-handed neutrinos gives
negligible contribution to the asymmetry.
We figure out general features of the washout effect.
It is shown that there is a novel parameter region where 
an effect that is negligible in most cases plays a critical role
and a sizable lepton asymmetry can survive against the 
washout process even in a strong washout region. 
\end{abstract}
\end{titlepage}
\newpage

\setcounter{page}{1}
\setlength{\parskip}{1.01ex} 

\renewcommand{\thefootnote}{\arabic{footnote}}
\setcounter{footnote}{0}
\section{Introduction}

Leptogenesis\cite{Lepto} is a simple mechanism to generate baryon number
asymmetry of the Universe.
The idea is that a lepton asymmetry produced at a high temperature is 
converted to the baryon asymmetry through the sphaleron 
interactions\cite{EWSph0} which conserve $B-L$ number but break 
$B+L$ number.
A simplest version of the leptogenesis is based on the 
seesaw mechanism\cite{Seesaw}
which can also explain the smallness of the neutrino 
masses by introducing heavy right-handed neutrinos (RHN) 
to the standard model.
In the seesaw models, the CP-violating and out-of-equilibrium decay of 
the RHN can produce the lepton asymmetry.

The seesaw mechanism is also easy to be implemented
in supersymmetric and/or grand unified theories (GUTs) which 
are most attractive candidates for the physics beyond the 
standard model. 
In such a class of model, the successful leptogenesis is 
considered as a mechanism to generate the baryon asymmetry of the 
Universe.
However, in many models especially in GUT models, 
the mass ($M_1$) of the lightest 
RHN ($N_1$) is too small to generate the observed baryon asymmetry 
by the $N_1$ decay and the asymmetry should be produced through 
another mechanism.

Recently it is pointed out that 
flavor effects give significant contributions
to the leptogenesis
\cite{Barbieri:1999ma,Nielsen:2002pc, Vives, FlavorLepto,NNRR}.
One of the interesting phenomena in the flavored leptogenesis
is that the primordial lepton asymmetry 
generated by the decay of the second lightest RHN ($N_2$), 
of the inflaton or so
can remain against the washout by the lightest
one\cite{Vives, EGNN}. This is an interesting possibility to give
enough baryon asymmetry even when the mass of the lightest RHN
is too small.
In such a scenario, the study of the washout effect by the lightest
RHN is very important. 

In this paper, we study the detail of this flavored washout effect 
due to the lightest RHN and we point that there is a novel parameter region 
where an effect that is negligible in most cases plays an important role. 
This effect is due to off-diagonal elements of the so-called A-matrix, and 
thus unique in the flavored leptogenesis. 
Most recently, the effects of the off-diagonal elements are studied 
numerically in the context that $N_1$ decay generates 
the lepton asymmetry\cite{Josse-Michaux:2007zj}.
Here, we adapt the analysis on the washout effect to the case 
where the asymmetry produced 
by the $N_2$ decay dominates the baryon asymmetry of the Universe, and show
that a sizable lepton asymmetry can remain against the washout process 
in a different way from those studied in Refs.\cite{Vives, EGNN}. 


In the section 2, we investigate the washout effect by the $N_1$ decay, 
assuming that there is primordial lepton asymmetry 
before the decay becomes relevant.
In the section 3, we will show examples that the primordial asymmetry is 
generated by the decay of the second lightest RHN.
The section 4 is devoted to summary and discussion.

\section{Flavor dependence of the Washout Effect}
\subsection{The seesaw mechanism and the leptogenesis}
In the seesaw mechanism, RHNs are introduced to the standard model,
\begin{equation}
\mathcal{L}=\mathcal{L}_{\text{SM}}+Y_{fj}h^*\bar{l}_fN_j
-\frac{M_i}{2}\bar{N}_iN_i^c\quad(i=1,2,3 \text{ and } f=e,\mu,\tau)\;,
\end{equation}
with $N_i$, $l_f$, and $h$ being RHNs, lepton doublets, and Higgs doublet
respectively.
Here we take the basis where the Yukawa matrix for charged leptons and 
the mass matrix of RHNs are diagonalized.
Integrating out the heavy RHNs and giving the VEV to Higgs, 
one can obtain the neutrino masses, $m_i$,
and Pontecorvo-Maki-Nakagawa-Sakata (PMNS) mixing matrix\cite{PMNS}, $U$, as
\begin{equation}
U^*\mathrm{diag}(m_1,m_2,m_3)U
=v^2Y\mathrm{diag}(M_1^{-1},M_2^{-1},M_3^{-1})Y^T\;.
\label{seesawrelation}
\end{equation}
where $v=174\GeV$ is the Higgs VEV. 
Supposing the reheating temperature after the inflation is enough high, 
the RHNs are produced through the interaction with the doublet leptons and 
the Higgs fields.
When the temperature decreases down to the mass of RHN, the production 
becomes inefficient and RHNs decay away. 
This out-of-equilibrium decay of the RHNs generates $B-L$ asymmetry
which is proportional to the CP violation in the decay, $\epsilon$, 
defined as
\begin{equation}
 \epsilon_i^f=\frac{\Gamma_{N_i\to l_fh}-\Gamma_{N_i\to \bar l_f\bar h}}
                   {\sum_f\PareS{\Gamma_{N_i\to l_fh}
                                +\Gamma_{N_i\to \bar l_f\bar h}}}.
\end{equation}
This asymmetry is converted to the baryon asymmetry through 
the electroweak sphaleron\cite{EWSph0}
processes.

\subsection{Boltzmann Equation}
\label{BoltzmannEq}

In order to evaluate the baryon asymmetry of the Universe,
the Boltzmann equation is used.
In this analysis, for simplicity, we omit the scattering effects which 
are considered to be subdominant. 
The decays and inverse decays, $N_1\leftrightarrow l_fh,\,\bar l_f\bar h$, 
are considered with rate $\gamma_D^f$.
With this simplification, the evolution of the asymmetry of 
$\Delta_f=B/3-L_f$ after the decoupling of the second lightest RHN 
is described by the following set of Boltzmann equations\cite{NNRR}:
\begin{eqnarray}
 &&\frac{Y_{N_1}}{dz}= -\frac{z}{sH(M_1)}\gamma_D
                 \PareS{\frac{Y_{N_1}}{Y_{N_1}^\eq}-1}, 
\label{BoltzmannEqN}\\
 &&\frac{dy_{\Delta_f}}{dz}= -\frac{z}{sH(M_1)}
                      \PareL{\gamma_D\epsilon_1^f
                             \PareS{\frac{Y_{N_1}}{Y_{N_1}^\eq}-1}
                            +\frac{\gamma_D^f}2
                             \PareS{\frac{y_{l_f}}{Y_{l_f}^\eq}
                                   +\frac{y_h}{Y_h^\eq}}
                      },
\label{BoltzmannEqB-L}
\end{eqnarray}
where $z=M_1/T$ and $\gamma_D=\sum_f\gamma_D^f$. 
The parameters $Y_X$ and $Y_X^{\text{eq}}$ indicate
the number density of the particle $X$ divided by the entropy
density $s=2\pi^2g_*^{\text{eff}}T^3/45$ 
and its value in equilibrium respectively,
and $y_X=Y_X-Y_{\bar X}$.
The parameter $g_*^{\text{eff}}\sim g_\SM^{\text{eff}}=106.75$ is the total effective number of the 
degrees of freedom (DOF) at the temperature around $M_1$.
With these definitions, we have
\begin{equation}
  Y_{N_1}^\eq=\frac34\frac{45\zeta(3)g_{N_1}}{4\pi^4g_*^{\text{eff}}}z^2K_2(z), \quad
  \frac{Y_{N_1}^\eq}{Y_{\rm massless}^\eq}=
  \half \frac{g_{N_1}}{g_{\rm massless}}z^2K_2(z)
  \Lm
  \begin{array}{lll}
   1 &\quad& {\rm for~fermion} \\
   3/4 &\quad& {\rm for~boson}
  \end{array}\RR
\label{Yeq}
\end{equation}
 where $\zeta(x)$ is the Riemann's zeta function and 
 $K_\nu(x)$ is the modified Bessel function.
Here, $g_X$ is (not effective) number of DOF of the particle $X$, 
 for example $g_{l_f}=g_{\bar l_f}=2$ and $g_{N_1}=2$. 

After neglecting the finite temperature effects such as the thermal masses 
and running of the couplings 
(for these effects, see Ref.\cite{FiniteTemp}) 
for simplicity, one can obtain $\gamma_D$ and the Hubble parameter $H(z)$ 
as 
\begin{equation}
 \gamma_D=sY_{N_1}^\eq\frac{K_1(z)}{K_2(z)}\Gamma_D, \quad
 H(T)=\sqrt{\frac{8\pi^3g_*^{\text{eff}}}{90}}\frac{T^2}{M_{pl}},
\end{equation}
 where $\Gamma_D=(Y^\dagger Y)_{11}M_1/(8\pi)$ 
 is the total decay width of $N_1$ and 
 $M_{pl}=1.22\times10^{19}\GeV$ is the Planck scale. 
Now, let us define the ``washout mass parameter'' $\tilde m_i^f$ and 
``equilibrium neutrino mass parameter'' $m_*$ as
\begin{equation}
 \tilde m_i^f = \frac{\abs{Y_{fi}}^2v^2}{M_i},\quad
 m_*=\frac{H(M_1)\tilde{m}_1}{\Gamma_D}
    =\sqrt{\frac{8\pi^3g_*^{\text{eff}}}{90}}\frac{8\pi v^2}{M_{pl}}=1.07\,\meV,
\label{def;mtilde}
\end{equation}
where $\tilde m_i=\sum_f\tilde m_i^f$. 
The partial decay width to a flavor $f$ 
is written as $\Gamma_D^f=\tilde m_1^f M_1^2/(8\pi v^2)$ and 
the total decay width is given by the sum of them 
$\Gamma_D=\sum\Gamma_D^f$.
Eventually, the Boltzmann equations are written as%
\begin{eqnarray}
 &&\frac{dY_{N_1}}{dz}= -z\frac{K_1(z)}{K_2(z)}
                 \frac{\tilde m_1}{m_*}
                 \PareS{Y_{N_1}-Y_{N_1}^\eq}, 
\label{BoltzmannEqN1}\\
 &&\frac{dy_{\Delta_f}}{dz}= -z\frac{K_1(z)}{K_2(z)}
                      \PareL{\epsilon_1^f\frac{\tilde m_1}{m_*}
                             \PareS{{Y_{N_1}}-{Y_{N_1}^\eq}}
                            +\frac14\frac{\tilde m_1^f}{m_*}
                             z^2K_2(z)
                             \PareS{2\frac{y_{l_f}}{g_{l_f}}
                                   +\frac32\frac{y_h}{g_h}}
                      }.
\label{BoltzmannEqB-L1}
\end{eqnarray}
The coefficient of $y_h$ is different from that in Ref.\cite{NNRR} 
by a factor $3/4$ which comes from the relative factor of the 
number density in the equilibrium of fermion/boson (\ref{Yeq}). 
Notice that, however, in the derivation of the Boltzmann equations 
(\ref{BoltzmannEqN}) and (\ref{BoltzmannEqB-L}), an approximation 
$f=1/\PareS{\exp\PareS{(E-\mu)/T}\pm1}\sim\exp\PareS{-(E-\mu)/T}$
is made. Within this approximation, the relative factor $3/4$ 
(and the factor $2$ in Eq.(\ref{mu2y})) disappears. 
In fact the right hand side of Eq.(\ref{BoltzmannEqB-L})
is originally written in terms of not 
$y_X/Y_X^{\text{eq}}$
but the chemical potentials of the particle $X$, $\mu_X$.
In literatures, these chemical potentials are replaced as
Eq.(\ref{BoltzmannEqB-L}) using the relation between 
$\mu_X$ and $y_X/Y_X^{\text{eq}}$ with the approximation.
If one would not use this replacement, 
a factor 1/2 appears instead of 3/4.
This difference of the factor does not affect the results a lot
in many cases and the term of $y_h$ itself is often neglected.
In our analysis, the contribution can affect the result significantly.
In the following we take basically the factor 3/4 for illustration.
It is straightforward to make analyses with a different factor.

An important point is that $y_{\Delta_f}$ is invariant under the standard 
evolution of the Universe and related to the present baryon asymmetry 
as $y_B=12/37\times\sum y_{\Delta_f}$\cite{EWSph}%
\footnote{
The factor $12/37$ can be somewhat different,
for instance $28/79$\cite{EWSph2}, 
depending on the timing of the freeze out of the electroweak sphaleron, 
but in any case, the value is approximately equal to $1/3$.
}
at the weak scale due to the electroweak sphaleron process.
Thus, we define ``baryon asymmetry'' by multiplying 
 the factor $12/37$ on $y_{\Delta_f}$ even at a higher temperature. 
Because this value is proportional to the present baryon to 
 photon ratio as $y_B=g_0^{\text{eff}}\pi^4/(45\zeta(3))\times \eta_B=7.04\eta_B$ 
 where $g_0^{\text{eff}}=43/11$ is the present total effective number of DOF, 
 successful leptogenesis scenario should predict
 the ``baryon asymmetry''
 $y_B^\obs=0.87\pm0.03\times10^{-10}$
 which comes from the observable
 $\eta_B^\obs=6.1\pm0.2\times10^{-10}$\cite{BAsymmObs}.

As mentioned in the introduction, in some models, 
the CP violation $\epsilon_1^f$ 
 is too small to produce enough lepton asymmetry.
In this case we can neglect the source term. 
Then, the evolution of $y_{\Delta_f}$ is controlled 
 by one equation as
\begin{equation}
 \frac{dy_{\Delta_f}}{dz}= -\frac14z^3K_1(z)
                     \frac{\tilde m_1^f}{m_*}
                     \PareS{2\frac{y_{l_f}}{g_{l_f}}
                            +\frac32\frac{y_h}{g_h}}
\end{equation}

Note that we assume that the asymmetries of the lepton doublet, $y_L$, and 
 of the Higgs fields, $y_H$, are much smaller than 1 and thus neglect 
 the higher terms because it is proportional to the $B-L$ asymmetry, 
 as shown below. 
These relations are forced by the fast (spectator) processes, such as 
 the sphaleron process, and depend on the temperature. 
For example at a high temperature, only the interactions mediated by 
the gauge and the top Yukawa coupling
are in the thermal equilibrium, while at a lower temperature 
weaker interactions come in it. 
For instance, let us concentrate on the range of the temperature where 
the interactions mediated by 
all the second and third generational Yukawa couplings are 
in the equilibrium but 
the first generational ones are not. 
This range is likely
the one in which we are interested, namely $T\sim M_1<10^9\GeV$. 
In this range, the weak sphaleron and the QCD sphaleron\cite{QCDSph}
are considered to occur fast enough.

These fast interactions make the following relations hold 
among the chemical potentials :
\[
\begin{array}{ll}
 \mu_{q_i}-\mu_{u_j}+\mu_H=0, &i=1,2,3\,\,\mbox{(due to the CKM mixing)}\\
 \mu_{q_i}-\mu_{d_j}-\mu_H=0, &j=2,3 \\
 \mu_{l_j}-\mu_{e_j}-\mu_H=0, \\
 \sum_i\PareS{3\mu_{q_i}+\mu_{l_i}}=0, & \mbox{EW sphaleron} \\
 \sum_i\PareS{2\mu_{q_i}-\mu_{u_i}-\mu_{d_i}}=0, \quad
   &\mbox{QCD sphaleron}.
\end{array}
\]
In addition to these relations, we impose the charge neutrality of the 
 Universe and assume the vanishing asymmetries for the right handed 
 leptons (and quarks if the QCD sphaleron is not considered) 
of the first generation:
\begin{eqnarray}
 &&\sum_i\PareS{\mu_{q_i}+2\mu_{u_i}-\mu_{d_i}-\mu_{l_i}-\mu_{e_j}}
   +2\mu_h=0, 
\label{Chemical-Charge}\\
 &&\mu_{e_1}=0, \\
 &&\mu_{u_1}=\mu_{d_1}(=0\quad\mbox{if no QCD sphaleron})
\end{eqnarray}
Notice that in the Eq.(\ref{Chemical-Charge}), the factor $2$ in front 
 of $\mu_h$ comes from the relative factor in the relation between 
 the asymmetry density and the chemical potential for massless particles :
\begin{equation}
 y_X
 = \frac{g_X\mu_X}{3s}T^2
    \Lm
    \begin{array}{lll}
     1/2 &\quad& {\rm for~fermion} \\
     1 &\quad& {\rm for~boson}
    \end{array}\RR.
\label{mu2y}
\end{equation}
Taking care of this factor 2, 
 we get similar relations among the asymmetries by replacing 
 $\mu_{\rm fermion}\to y_{\rm fermion}/g_{\rm fermion}$, $\mu_h\to 2y_h/g_h$
 and 
 $\sum_i\PareS{\frac13\times3\times\PareS{2\mu_{q_i}+\mu_{u_i}+\mu_{d_i}}}/3 
 -\PareS{2\mu_{l_f}+\mu_{e_f}} \to y_{\Delta_f}$.
By solving these relations, we find the expression of $y_{l_f}/g_{l_f}$ and 
 $y_h/g_h$ 
 in terms of $y_{\Delta_i}$ as 
\begin{equation}
  \frac{y_{l_f}}{g_{l_f}}={C_l}_{ff'}y_{\Delta_{f'}}, \qquad 
  \frac34\frac{y_h}{g_h}=\frac34{C_h}_fy_{\Delta_f}, 
\end{equation}
with 
\begin{equation}
  C_l=
  \begin{pmatrix}
    -109/253  &   25/506  &   25/506  \\
      29/1012 & -493/1518 &   13/1518 \\
      29/1012 &   13/1518 & -493/1518
  \end{pmatrix}, \qquad
  C_h=
  \begin{pmatrix}
     -53/506 \\
     -37/253 \\
     -37/253  
  \end{pmatrix}
\label{Cw/oQCDS}
\end{equation}
if we do not consider the QCD sphaleron and with 
\begin{equation}
  C_l=
  \begin{pmatrix}
    -151/358 &   10/179 &   10/179 \\
      25/716 & -172/537 &    7/537 \\
      25/716 &    7/537 & -172/537
  \end{pmatrix}, \qquad
  C_h=
  \begin{pmatrix}
     -37/358 \\
     -26/179 \\
     -26/179  
  \end{pmatrix}
\label{CwQCDS}
\end{equation}
if we take into account it.
In the following, we examine only the latter case because there are no 
 qualitative difference.

\begin{figure}
\begin{center}
  \includegraphics[scale=2.0]{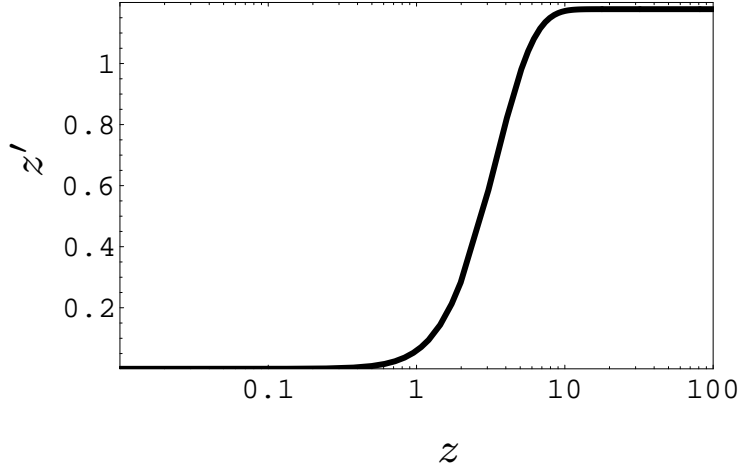}
\caption{
The new variable $z'$ as a function of $z$.
}
\label{fig:sVSz}
\end{center}\end{figure}

From these expressions, we have  the following Boltzmann equation:
\begin{equation}
 \frac{dy_{\Delta_f}}{dz} = 
   -\frac{z^3}4 K_1(z)\frac{\tilde m_1^i}{m_*}
    A_{ff'} y_{\Delta_{f'}},
\end{equation}
with 
\begin{equation}
 A_{ff'}=
  \begin{pmatrix}
    715/716 &  19/179 &  19/179 \\
     61/716 & 461/537 & 103/537 \\
     61/716 & 103/537 & 461/527
  \end{pmatrix}
  =
  \begin{pmatrix}
     1.00  & 0.11 & 0.11 \\
     0.085 & 0.86 & 0.19 \\
     0.085 & 0.19 & 0.86
  \end{pmatrix},
\end{equation}
where the summation over $f'=e,\mu,\tau$ should be understood.
In order to analyse this equation, it is convenient to change the 
 variable from $z$ to $z'$ that satisfy $\frac{d z'}{dz}=z^3 K_1(z)/4$ so that 
\begin{equation}
 \deriv{y_{\Delta_f}}{z'} = - \frac{\tilde m_1^f}{m_*} A_{ff'} y_{\Delta_{f'}}.
\label{BoltzmannEqF}
\end{equation}
The range of $z'$ is from $z'(z=0)=0$ to $z'_\infty=z'(z=\infty)=3\pi/8=1.18$.
The relation between them is shown in the Fig.\ref{fig:sVSz}.
This figure shows the washout occurs mostly in the temperature range 
$M_1/10 \lesssim T \lesssim 10M_1$.

For comparison, the Boltzmann equation for the usual one-flavor 
approximation, which is in reality valid only when the temperature 
is high enough that even the processes mediated by the tau Yukawa coupling
are out of equilibrium, 
is given as\footnote{
Here we neglect the Higgs contribution in order to compare 
our analysis with those in literatures, though
it is considered in the next section.
} 
\begin{equation}
 \deriv{y_{\Delta}}{z'} = - \frac{\tilde m_1}{m_*} y_{\Delta},
\label{BoltzmannEqNF}
\end{equation}
where $y_{\Delta}=\sum y_{\Delta_f}$ and $\tilde m_1=\sum \tilde m_1^f$.

\subsection{Solutions}
\label{solutions}

Roughly speaking, the matrix $A_{ff'}$ is close to a diagonal one. 
And thus, we can find an approximate solution by a perturbation 
with respect to rather small off-diagonal elements. 
Neglecting the off-diagonal elements, 
each initial asymmetry $y_{\Delta_f}^0$ is 
exponentially washed out.
The evolution of the asymmetry is given by 
\begin{equation}
  y_{\Delta_f}^{(0)}(z')
 =\exp\PareS{-\frac{\tilde m_1^f}{m_*}A_{ff}z'}y_{\Delta_f}^0.
\label{yi0th}
\end{equation}

When the off-diagonal elements are switched on,
the asymmetry follows
\begin{equation}
  \deriv{y_{\Delta_f}^{(1)}}{z'}
 =- \frac{\tilde m_1^f}{m_*} A_{ff} y_{\Delta_f}^{(1)}
  - \sum_{f'\neq f}\frac{\tilde m_1^f}{m_*} A_{ff'} y_{\Delta_{f'}}^{(0)},
\end{equation}
up to the next leading order of the off-diagonal elements.
Inserting the leading order solution Eq.(\ref{yi0th}), we find 
\begin{eqnarray}
  y_{\Delta_F}^{(1)}(z')
 &=& \sum_{f'\neq f} \frac{\frac{\tilde m_1^f}{m_*} A_{ff'}}
                       {\frac{\tilde m_1^{f'}}{m_*} A_{f'f'}
                       -\frac{\tilde m_1^f}{m_*} A_{ff}}
   \PareS{
     \exp\PareS{-\frac{\tilde m_1^{f'}}{m_*}A_{f'f'}z'}
    -\exp\PareS{-\frac{\tilde m_1^f}{m_*}A_{ff}z'}
   }y_{\Delta_{f'}}^0 
\label{yi1st}
.
\end{eqnarray}
This expression shows that even if the initial asymmetry of a certain flavor 
 is zero, the asymmetry is generated from those of the others 
 although it is suppressed by a small off-diagonal element of $A_{ij}$. 
For most cases, this order already gives good approximation for the 
 final value of the total asymmetry.

\subsubsection*{(1) $\tilde{m}_1^f\lesssim m_*$
($f=e,\mu,\tau$)}

In this case, one may expect 
 the washout effect is small and thus the flavor effect 
 can not play an important role.
However, even in this case, the final total asymmetry can be 
 2 times larger than the one-flavor approximation 
 (See (1) in Fig.\ref{fig:Washout}).

\subsubsection*{(2) $\tilde{m}_1^f \gnsim m_*$ and $\tilde{m}_1^{f'}\lesssim m_*$ ($f\neq f'$)}

%

In this case, the summation $\tilde m_1=\sum\tilde m_1^f$ is dominated
by $\tilde{m}_1^{f'}$ and is
larger than $m_*$. It is called strong washout region.
As shown below, however, if we take account of the flavor effect 
the washout effect is drastically changed.
The effect depends strongly on the flavor structure of the initial asymmetry.

\begin{figure}
\begin{center}
\begin{tabular}{cc}
\includegraphics[scale=1.3]{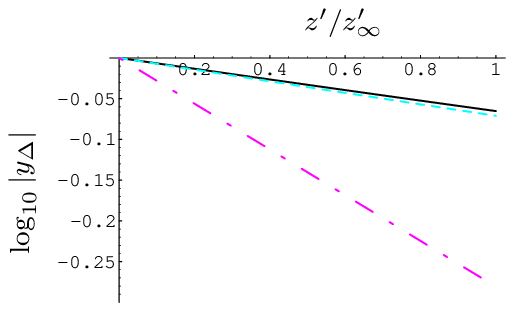}
&
\includegraphics[scale=1.3]{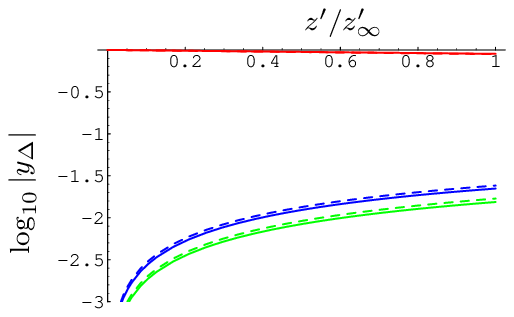} \\
\multicolumn{2}{c}{\vspace{-5mm} (1) $\PareM{\tilde m_1^e,\tilde m_1^\mu,\tilde m_1^\tau}
=\PareM{0.1,0.2,0.3}\meV$\;,\quad
$\{y_{\Delta_e}^0,y_{\Delta_\mu}^0,y_{\Delta_\tau}^0\}=\PareM{1,0,0}$
}\\[5mm]
\includegraphics[scale=1.3]{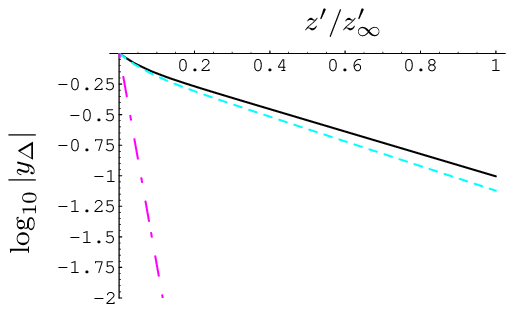}
&
\includegraphics[scale=1.3]{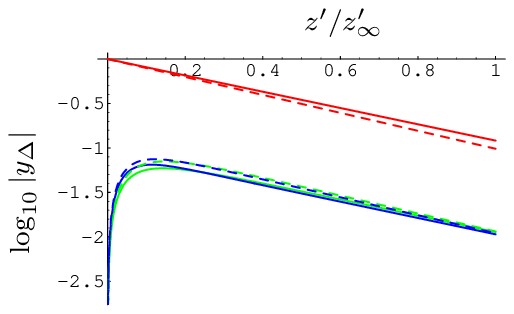} \\
\multicolumn{2}{c}{\vspace{-5mm} (2a) $\PareM{\tilde m_1^e,\tilde m_1^\mu,\tilde m_1^\tau}
=\PareM{2, 15,20}\meV$\;,\quad
$\{y_{\Delta_e}^0,y_{\Delta_\mu}^0,y_{\Delta_\tau}^0\}=\PareM{1,0,0}$
}\\[5mm]
\includegraphics[scale=1.3]{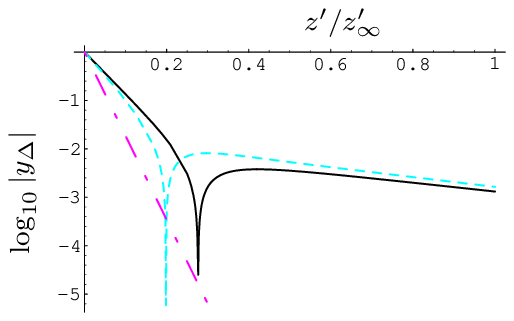}
&
\includegraphics[scale=1.3]{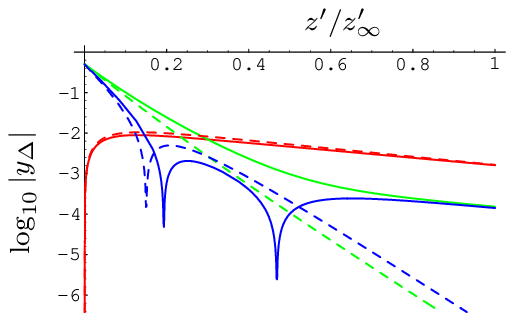} \\
\multicolumn{2}{c}{\vspace{-5mm} (2b) $\PareM{\tilde m_1^e,\tilde m_1^\mu,\tilde m_1^\tau}
=\PareM{2, 15,20}\meV$\;,\quad
$\{y_{\Delta_e}^0,y_{\Delta_\mu}^0,y_{\Delta_\tau}^0\}=\PareM{0,0.5,0.5}$
}\\[5mm]
\end{tabular}
\caption{
The evolutions of the $B/3-L_f$ asymmetries.
The horizontal line is $z'/z'_\infty$, and the vertical line is 
$\log_{10}\abs{y_{\Delta}}$.
In the left figures, black (solid), light blue (broken) 
and purple (dot-dashed) lines show the total 
$B-L$ asymmetry calculated by the full Boltzmann equation 
(\ref{BoltzmannEqF}), by the approximation formula (\ref{yi1st}) and 
the one-flavor approximation (\ref{BoltzmannEqNF}), respectively.
In the right figures, red, green and blue lines respectively show 
 the $\Delta_e$, $\Delta_\mu$ and $\Delta_\tau$, and the solid and broken 
 lines corresponds (\ref{BoltzmannEqF}) and (\ref{yi1st}), respectively.
}
\label{fig:Washout}
\end{center}\end{figure}

In the following, let us take 
$\tilde m_1^e \lesssim m_* \lnsim \tilde m_1^a\,(a=\mu,\tau)$ 
as a representative example for clarity. 
It is straightforward to apply this analysis to the other cases.
We consider two typical sets of the initial asymmetries:
\begin{itemize}
\item[(2a)] $y_{\Delta_e}^0 \gtrsim y_{\Delta_a}^0$\\
In this case, the washout effect of $y_{\Delta_e}$ is controlled by 
$\tilde m_1^e$ which 
is much smaller than $\tilde m_1$, while $y_{\Delta_a}$ are generated 
due to the small off-diagonal elements of $A_{ea}$, 
with the opposite sign.
Because of the large $\tilde m_1^a$, these generated $y_{\Delta_a}$ 
are washed out strongly and can not become comparable with $y_{\Delta_e}$.
Thus, in terms of the perturbation, the leading order approximation is 
sufficient.

Note that even in this case, the washout factor 
$y_{\Delta}(z'_\infty)/y_{\Delta}^0\sim\exp\PareS{\tilde m_1^eA_{ee}/m_*}$ 
is quite different (much larger) than the one-flavor approximation, 
$\exp\PareS{\tilde m_1/m_*}$ (See (2a) in Fig.\ref{fig:Washout}). 
This is the case even when $\tilde m_1^e$ is not so much smaller than 
the others, due to the exponential washout factor\cite{Vives}. 
\item[(2b)] $y_{\Delta_e}^0\ll y_{\Delta_{\mu}}^0$ and/or $y_{\Delta_{\tau}}$\\
The asymmetry $y_{\Delta_a}$ decreases rapidly, while the
asymmetry $y_{\Delta_e}$  produced due to the
off-diagonal elements is washed out much more slowly. 
This means that at some point $y_{\Delta_e}$ becomes dominant. 
Once it becomes dominant, the following evolution is similar to the one 
in the case (2a).
Thus, the washout factor is controlled basically by the small $\tilde m_1^e$ 
rather than $\tilde m_1$ or $\tilde m_1^a$ 
(See (2b) in Fig.\ref{fig:Washout}). 
Interestingly in this case, the sign of the total $B-L$ asymmetry changes 
 through the washout.

Note that the approximation at the NLO is quite bad for $y_{\Delta_a}$ 
because the secondary conversion from $y_{\Delta_e}$, which is generated 
by the NNLO effect, is important. 
Nevertheless, the approximation for the total asymmetry is rather good 
because these are small as in the case (2a).
\end{itemize}

\subsubsection*{(3) $\tilde m_1^f \gnsim m_*$ $(f=e,\mu,\tau)$}

In this case, all the asymmetries in each flavors are strongly washed out. 
Thus, it is hard that the observed value remains after the washout, 
 as far as $M_1<10^9\GeV$%
\footnote{
If we consider models with $M_1\gg10^9\GeV$ where the 
 muon Yukawa interaction 
 is out of equilibrium, the asymmetry along with the direction in the flavor 
 space that is orthogonal both to the direction determined by $N_1$ Yukawa 
 coupling and $\tau$ direction is free from the washout, 
 even in this case\cite{EGNN}. 
}.

From the above considerations, it is clear that the washout factor is 
 basically controlled by the smallest washout mass parameter. 
This is in great contrast to the non-flavored case, where the 
 factor is controlled basically by the largest washout mass parameter.
Interestingly, this is also true even for the case that the initial asymmetry 
 of the flavor with smallest washout mass parameter is tiny. 
For this case, the effect of the off-diagonal elements, 
 which is usually negligible, is crucial.

\subsection{Fixed point}

From Eq.~(\ref{BoltzmannEqF}), one can obtain a coupled
equations for $y_{\Delta_f}/y_{\Delta_{\tau}}\,(f=e,\mu)$ as
\begin{equation}
\frac{d}{dz'}\left(\frac{y_{\Delta_f}}{y_{\Delta_{\tau}}}\right)
=\sum_{f'=e,\mu,\tau}\frac{\tilde{m}_1^{\tau}}{m_*}
\left[
\frac{\tilde{m}_1^f}{\tilde{m}_1^{\tau}}A_{ff'}
-A_{\tau f}\left(\frac{y_{\Delta_f}}{y_{\Delta_{\tau}}}\right)
\right]\left(\frac{y_{\Delta_{f'}}}{y_{\Delta_{\tau}}}\right)\;.
\end{equation}
This couple of equations have fixed points in the space of
$(y_{\Delta_e}/y_{\Delta_{\tau}}, y_{\Delta_{\mu}}/y_{\Delta_{\tau}})$,
which is determined by solving the equations
\begin{align}
\sum_{f'=e,\mu,\tau}\left[
\frac{\tilde{m}_1^f}{\tilde{m}_1^{\tau}}A_{ff'}
-A_{\tau f'}\left(\frac{y_{\Delta_f}}{y_{\Delta_{\tau}}}\right)
\right]\left(\frac{y_{\Delta_{f'}}}{y_{\Delta_{\tau}}}\right)
=0\;.
\end{align}
Once the flow of the solution reaches close to 
a fixed point at $z'=z'_{\text{fp}}$, the set of ratios
$(y_{\Delta_e}/y_{\Delta_{\tau}}, y_{\Delta_{\mu}}/y_{\Delta_{\tau}})$ 
becomes invariant and the Boltzmann equations can be rewritten 
as
\begin{align}
\frac{dy_{\Delta_f}}{dz'}=-\frac{\tilde{m}_1^\text{fp}}{m_*}
y_{\Delta_f}\;,\quad z'\geq z'_{\text{fp}}\;.
\end{align}
This means that the asymmetries of the all flavor are washed out
with the universal washout mass parameter $\tilde{m}_1^{\text{fp}}$
which corresponds to one of the eigenvalues
of matrix $B_{ff'}\equiv \tilde{m}_1^fA_{ff'}$.
There are three possible points in the case of three effective 
flavor numbers. However two of them are unstable fixed points 
and only one point is attractive. The attractive one most likely
corresponds to the smallest eigenvalue which is smaller than
the smallest $\tilde{m}_1^f$.
Around the attractive fixed point, the asymmetry in the flavor with
the smallest washout mass parameter, $y_{\Delta_{f_1}}$
dominates the total asymmetry.
This can be understood as follows.
The asymmetry in the flavor with larger washout mass parameter,
$y_{\Delta_{f_2}}$ decrease
more quickly as discussed in the case (2b) in the Sec.~\ref{solutions}. 
When $y_{\Delta_{f_2}}$ becomes 
the order of $A_{f_1f_2}y_{\Delta_{f_1}}$, 
$y_{\Delta_{f_2}}$ evolves similar to $y_{\Delta_{f_1}}$
because the transportation from $y_{\Delta_{f_1}}$ becomes to control the 
evolution.
Then the washout of $y_{\Delta_{f_1}}$ becomes weaker due to 
the transportation from $y_{\Delta_{f_2}}$.

When the initial condition is too far from the fixed point and
$\tilde{m}_1\ll m_*$, the washout term decouples from the system 
before the solution flows into the fixed point\footnote{
Notice that the $z'$ takes the value in the range of $0\leq z'\leq 3\pi/8$
with respect to $0\leq z\leq \infty$.
}.

One can see similar phenomena also in the case where only $N_1$ 
decay produces the asymmetries and they are washed out by 
the $N_1$ (inverse) decay.
In fact effects of off-diagonal elements of $A$ matrix 
in such a case are discussed in Ref.\cite{Josse-Michaux:2007zj} 
and the asymmetry shown there behaves as discussed in the above.

\section{Asymmetries by the Second Lightest RHN Decay}

In this section, we investigate the possibility that the initial asymmetries 
are generated via the second lightest RHN decay.

For simplicity, we restrict ourselves to the case that the mass of the 
second lightest RHN is larger than $10^{12}\GeV$. 
This case is qualitatively discussed in Ref.\cite{EGNN}.
For this mass range, the fast interactions that are in the equilibrium when 
the RHN decays are only the interactions mediated by 
the top Yukawa coupling and QCD sphalerons. 
This means that the fast interactions can not distinguish all the 
generations of the lepton doublets, and thus two liner combinations of 
the three doublets that do not interact with the RHN are never produced.
Namely, only $l_\parallel\propto Y_{\tau2}l_\tau+Y_{\mu2}l_\mu+Y_{e2}l_e$ 
are produced. 
Then, the relevant Boltzmann equations are for one flavor system, 
which is given by Eqs.(\ref{BoltzmannEqN1}) and (\ref{BoltzmannEqB-L1}) 
by replacing all the index $1$ to $2$ (including $z\to M_2/T$) and 
 suppressing the flavor indexes. 
With the definitions given in the section \ref{BoltzmannEq}, 
 we have
\begin{eqnarray}
 &&\frac{d Y_{N_2}}{dz}=\frac{\tilde m_2}{m_*} z \frac{K_1(z)}{K_2(z)}
                \Ls Y_{N_2}-Y_{N_2}^\eq \Rs \\
 &&\frac{d y_{\Delta}}{dz}=\epsilon_2
                     \frac{\tilde m_2}{m_*} z \frac{K_1(z)}{K_2(z)}
                     \Ls Y_{N_2}-Y_{N_2}^\eq \Rs
                    -\frac{z^3}4 K_1(z)\frac{\tilde m_2}{m_*}
                     A y_{\Delta}.
\end{eqnarray}
The A factor for this case is calculated in a similar way to the discussion 
 in the section \ref{BoltzmannEq} as 
\begin{equation}
 A=2C_l+\frac32C_h=\frac{67}{46}
\end{equation}
It is possible, of course, that we solve these set of equations numerically 
 to evaluate the $B-L$ asymmetry produced by the decay of 
 the second lightest RHN.
In this article, however, we use the following approximation formula proposed 
 in Ref.\cite{Petcov}, which includes the effects of the scatterings, 
 to evaluate the ``baryon asymmetry'':
\begin{equation}
 y_B\sim-\frac{12}{37g_*^{\text{eff}}}
              \epsilon_2 \eta\PareS{A\tilde m_2}
\label{AppFormula}
\end{equation}
with
\begin{equation}
 \eta(x)=\PareS{\PareS{\frac{x}{8.25\,\meV}}^{-1}
               +\PareS{\frac{0.2\,\meV}{x}}^{-1.16}
               }^{-1}.
\end{equation}

In any case, these equations are controlled by the parameters $\tilde m_2$ 
 and $\epsilon_2$ which are determined by the mass spectrum of the RHNs 
 $M_i$ and the neutrino Yukawa coupling $Y_{fi}$.
To be more concrete, they are given by sums of (\ref{def;mtilde}) and  
\begin{eqnarray}
 &&\epsilon_2^f=\frac1{8\pi}
  \frac1{\PareS{Y^\dagger Y}_{22}}
  {\rm Im}\sum_{i\neq 2}Y^*_{f2}Y_{fi}
                \PareS{\PareS{Y^\dagger Y}_{2i}f\PareS{\frac{M_i^2}{M_2^2}}
                      +\PareS{Y^\dagger Y}_{i2}g\PareS{\frac{M_i^2}{M_2^2}}}.
\end{eqnarray}
Here both the diagrams of the vertex correction and 
 of the self-energy correction are implemented in each function as
\begin{eqnarray}
 && f(x)=-\frac{\sqrt{x}}{x-1}
         +\sqrt{x}\PareS{1-\PareS{1+x}\ln\PareS{\frac{1+x}{x}}}, \\
 && g(x)=-\frac{1}{x-1}.
\end{eqnarray}
In this way, fixing $M_i$ and $Y_{fi}$, all the parameters in the Boltzmann 
 equations are determined, and we can calculate the $B-L$ asymmetry 
 $y_{\Delta}$ just after the second lightest RHN decouples. 
This asymmetry is washed out when the lightest RHN starts decaying. 
In this period, the $\tau$ and $\mu$ Yukawa couplings enter into 
 the equilibrium, and thus the fast interactions distinguish 
 all the three flavors. 
Therefore, we should divide $y_{\Delta}$ into $y_{\Delta_\tau}$, 
 $y_{\Delta_\mu}$ and $y_{\Delta_e}$, which follow the relation 
 $y_{\Delta_\tau}:y_{\Delta_\mu}:y_{\Delta_e}
  =\tilde m_2^\tau:\tilde m_2^\mu:\tilde m_2^e$ 
 according to the probabilistic interpretation.
This set of asymmetries $y_{\Delta_f},\,(f=e,\mu,\tau)$ gives 
the initial condition of the analysis given in the section 
\ref{solutions}.

The neutrino Yukawa couplings $Y$ should be related to the low energy 
neutrino parameters through the seesaw relation, Eq.(\ref{seesawrelation}).
In order to represent the solution of this relation,
we adopt the following famous parameterization\cite{CASAS}, 
\begin{equation}
 Y_{fi}= \PareS{U^*}_{fj} \sqrt{m_j}R_{ji}\sqrt{M_i}/v. 
\end{equation}
Here $U$ is written as the product of a CKM-like mixing matrix $V$ 
which includes three mixing angles and one CP phase\footnote{
As a parametrization of $V$, we adopt the Chau--Keung parametrization 
(PDG parametrization) \cite{PDGparametrization}.
}
and a phase matrix with two Majorana phases 
$P=\diag\PareS{1,\exp\PareS{i\alpha_{21}/2},\exp\PareS{i\alpha_{31}/2}}$
: $U=V P$
and
$R$ is a complex orthogonal matrix which can be decomposed as 
$R=e^{i\omega_{23}\lambda_7}e^{i\omega_{13}\lambda_5}
e^{i\omega_{12}\lambda_2}$
where $\lambda_i$ are Gell-Mann matrices and $\omega_{ij}$ are
complex parameters.
For simplicity, in this article, we use the following set of the parameters 
for the light neutrino sector as 
$m_i=\PareM{0,9,50}\meV$, 
$\PareM{s12^2,s23^2,s13,\delta,\alpha_{21},\alpha_{31}}=
 \PareM{0.3,0.5,0,0,0,0}$ for the PMNS matrix, and 
Majorana masses $M_i=\PareM{10^7,10^{13},10^{14}}\GeV$ for the RHNs.

As representative examples, let us consider the following sets:
\begin{eqnarray}
 &{\rm (I)}& :\quad 
\PareM{\omega_{12},\omega_{23},\omega_{13}}=
\PareM{30^\circ,i5^\circ,-1^\circ}
\nn\\
 &{\rm (I\!I a)}& :\quad 
\PareM{\omega_{12},\omega_{23},\omega_{13}}=
\PareM{-88^\circ,(60+i3)^\circ,3^\circ}
\label{Examples}\\
 &{\rm (I\!I b)}& :\quad 
\PareM{\omega_{12},\omega_{23},\omega_{13}}=
\PareM{(-85+i4)^\circ,(50+i20)^\circ,-5.5^\circ} 
\nn
\end{eqnarray}

For instance, for the example (I), 
 we find 
\begin{equation}
 \tilde m=
  \begin{pmatrix}
    0.68 & 2.04 & 0.02 \\
    0.71 & 2.66 & 25.2 \\
    0.98 & 2.39 & 25.2
  \end{pmatrix}
  \meV
 ,\quad
 \epsilon=
  \begin{pmatrix}
    10^{-7} & -0.27 & -0.02 \\
    10^{-4} & -95.0 & 11.57 \\
    10^{-4} &  84.3 & -11.28
  \end{pmatrix}
  \times10^{-6}.
\end{equation}
These show 
 $\PareM{\tilde m_1^e,\tilde m_1^\mu,\tilde m_1^\tau}
  =\PareM{0.68,0.71,0.98}\meV \lesssim m_*$ and 
 $\PareM{\epsilon_1^e,\epsilon_1^\mu,\epsilon_1^\tau}
  =\PareM{10^{-13},10^{-11},10^{-11}}$ are negligibly small. 
Thus, this is an example of the case (1) in the section \ref{solutions}.
Using the approximation (\ref{AppFormula}), we see the ``baryon 
asymmetry'' generated by the $N_2$ decay is $Y_B^0=3.36\times10^{-10}$. 
When the temperature decrease to around $M_1$, $N_1$ starts decaying, 
and the asymmetry is washed out in the way investigated in the last section.
As mentioned above, in this period, the fast interactions distinguish all 
the flavor, and the asymmetry should be divided as 
 $\PareM{{y_{B_e}^0},{y_{B_{\mu}}^0},{y_{B_{\tau}}^0}}
 =\PareM{0.97,\,1.26,\,1.13}\times10^{-10}$.
After the washout, a total asymmetry $y_B=1.26\times10^{-10}$ remains.

In a similar way, (\II a) and (\II b) are examples of the cases 
 (2a) and (2b), respectively. 
Their results are listed in the Table \ref{Results}.

\begin{table}
\begin{center}
\begin{tabular}{cccc}
\hline
 & (I) & (\II a) & (\II b) 
\\\hline
 $\tilde m\,(\meV)$ 
 &
  $\begin{pmatrix}
    0.68 & 2.04 & 0.02 \\
    0.71 & 2.66 & 25.2 \\
    0.98 & 2.39 & 25.2
  \end{pmatrix}$
 &
  $\begin{pmatrix}
    0.68 & 0.01 & 2.03 \\
    12.0 & 0.01 & 16.3 \\
    27.3 & 0.02 & 1.00
  \end{pmatrix}$
 &
  $\begin{pmatrix}
    1.46 & 10^{-5} &  1.91 \\
   10.43 & 0.421   & 24.26 \\
   27.91 & 0.428   &  6.95
  \end{pmatrix}$
\\
 $\epsilon\,(10^{-6})$
 &
  $\begin{pmatrix}
    10^{-7} & -0.27 & -0.02 \\
    10^{-4} & -95.0 & 11.57 \\
    10^{-4} &  84.3 & -11.28
  \end{pmatrix}$
 &
  $\begin{pmatrix}
    10^{-6} & -8.97  & 0.01 \\
    10^{-5} & -31.73 & 0.02 \\
    10^{-5} & -5.02  &-0.02
  \end{pmatrix}$
 &
  $\begin{pmatrix}
    10^{-6} & -0.52 & 10^{-3} \\
    10^{-4} & 341   & -2.70   \\
    10^{-4} & 185   &  0.335
  \end{pmatrix}$
\\
 $y_B^0\,(10^{-10})$ 
 & $\PareM{0.97,\,1.26,\,1.13}$
 & $\PareM{2.53,\,2.09,\,4.28}$
 & $-\PareM{0.017,\,521,\,530}$
\\
 $y_B\,(10^{-10})$ 
 & $1.26$
 & $1.02$
 & $2.00$
\\\hline
\end{tabular}
\caption{
Results of $\tilde m$, $\epsilon$, $Y_B^0$ and $Y_B$ for the three examples 
 in (\ref{Examples}).
}
\end{center}
\label{Results}
\end{table}

\section{Summary and Discussions}

In this article, we investigate the washout effect due to the $N_1$ 
(inverse) decay, assuming non-vanishing initial lepton asymmetry 
and negligible lepton asymmetry production in $N_1$ decay. 
We show that there is a novel parameter region in addition to those 
studied in Refs.\cite{Vives, EGNN}.
There, off-diagonal elements of the $A$-matrix, which are often omitted, 
play a critical role. 
This region is where some of $\tilde m_1^f$ is comparable to 
or smaller than $m_*$, the others are larger than it,
and the initial asymmetries on the flavors with small $\tilde m_1^f$ 
are tiny. 
In this case, if we would omit the off diagonal elements as usual, 
any initial asymmetries on the flavors with large $\tilde m_1^f$ 
were strongly washed out. 
In fact, the off diagonal elements transform the asymmetries 
from those with large $\tilde m_1^f$ to those with small ones. 
Once transformed, such asymmetries are weakly washed out, and thus 
a sizable total asymmetry may survive.

For completeness, we examine the possibility that the initial asymmetry 
is generated by the $N_2$ decay within the thermal leptogenesis scenario.
We show some concrete examples for each class discussed in the above 
analysis.

Finally, let us make a comment on an ambiguity of the Boltzmann equations, 
especially on the factor in front of $y_h$ in Eq.(\ref{BoltzmannEqB-L1}).
As briefly discussed below the equation, 
an approximation is used in the derivation of the Boltzmann equations 
(\ref{BoltzmannEqN}), (\ref{BoltzmannEqB-L}), and it brings the ambiguity.
Because the contribution of this term ($y_h$) is relatively small, 
as seen from (\ref{Cw/oQCDS}) and (\ref{CwQCDS}), 
this ambiguity does not affect results so much, $\order{10\%}$. 
In the case (2b), however, the off diagonal element of $A$-matrix, which 
is of the same order as the $y_h$ contribution, is critical. 
In addition, a cancellation occurs between $y_h$ contribution and that of  
$y_l$ in the off diagonal element. 
Thus, the result is largely changed due to the ambiguity.
In fact, if we take a factor $1/2$ instead of $3/4$ as a possible choice,
the final baryon asymmetry is reduced to $y_B=0.83\times10^{-10}$ 
with the same parameters as (\II b) in (\ref{Examples}).
Thus, it is important to make a closer look on the Boltzmann equations 
before discussing this novel effect quantitatively.

\vspace{1cm}
{\bf Acknowledgements.}  We would like to thank
S.~T~.Petcov and Y.~Takanishi for useful discussions.
This work was supported in part by the Italian MIUR and 
INFN under the programs ``Fisica Astroparticellare''.

\end{document}